\documentclass[12pt]{article}
\usepackage{amsmath,amsfonts,amssymb,a4,color,graphics,epsf}
\usepackage[latin1]{inputenc}
\usepackage{verbatim}
\usepackage{times}

\def\bR {{\mathbb{R}}}
\def\bN {{\mathbb{N}}}

\def\bZ {{\mathbb{Z}}}

\def\pa{\partial}

\def\Di {\displaystyle}

\makeatletter
\@addtoreset{equation}{section}

\makeatother

\newtheorem{theorem}{Theorem}[section]
\newtheorem{lemma}[theorem]{Lemma}
\newtheorem{proposition}[theorem]{Proposition}

\newtheorem{corollary}[theorem]{Corollary}
\newtheorem{example}[theorem]{Example}

\newtheorem{remark}[theorem]{Remark}
\newenvironment{demo}{\noindent {\it Proof.--}
      \begin{quotation}\noindent}{\end{quotation}\hfill$\square $}

\begin{document}

\bibliographystyle{plain}

\title{ Eigenvalue bounds  for radial magnetic bottles  on the disk}

\author{
 Fran\c{c}oise Truc\footnote{Institut Fourier, 
francoise.truc@ujf-grenoble.fr
Unit{\'e} mixte
 de recherche CNRS-UJF 5582,
 BP 74, 38402-Saint Martin d'H\`eres Cedex (France)}}

\maketitle

\begin{abstract}
We consider  a 
 Schr\"odinger operator $H_A^D$ with a non-vanishing 
radial magnetic field 
$B=dA$ and Dirichlet boundary conditions on the unit disk. 
 We assume growth conditions on $B$
 near the boundary which
guarantee in particular the compactness of the resolvent of this operator.
 Under some assumptions on an additional radial potential $V$ the operator $H_A^D - V$ has a 
discrete negative 
spectrum and we obtain an upper bound of the number of negative eigenvalues.
 As a consequence we get an  upper bound of the
number  of eigenvalues of $H_A^D$ 
smaller than any positive value $\lambda$, which involves the minimum of $B$
 and the square of the $L^2$-norm 
of  $A(r)/r$, where $A(r)$ is the specific magnetic potential defined as the flux of the magnetic field through the disk of radius $r$ centered in the origin.
\end{abstract}

\section{ Introduction}
Let us consider a particle  in a  domain $\Omega$ in $\bR^2 $
in the presence of a  magnetic field $B$. We define the 2-dimensional
magnetic Laplacian associated to this particle as follows:\\
Let $A$ be a  magnetic potential associated to $B$ ; it means that $A$ 
is a  smooth real one-form   on
 $\Omega \subset \bR^2$, given by  $A=\sum_{j=1}^2  a_j dx_j$,
 and that the
  magnetic field $B$ is  the two-form $B=dA$.
We have $B(x)= {\bf b} (x) dx_1\wedge dx_2$ 
 with $ {\bf b}(x) = \partial_1 a_2(x)-\partial_2 a_1(x)\ .$
The  magnetic connection $\nabla =(\nabla _j)$ is the differential 
operator defined
by
\[ \nabla _j=
\frac{\pa }{\pa x_j } -i a_j  ~.\]
The 2-dimensional {\rm magnetic Schr\"odinger} operator $H_A$ is defined
by
 \[H_A =-\sum_{j=1}^2 \nabla _j^2 ~.\]
The  magnetic Dirichlet integral  $h_A=\langle H_A .| . \rangle $
is given, for $u\in C_0^\infty (\Omega )$,
by
\begin{equation}\label{quadra}
 h_A (u)= \int _\Omega \sum_{j=1}^2 |\nabla _j u|^2 |dx| ~.
\end{equation}
From the previous definitions and the fact that the formal adjoint
of $\nabla _j$ is $-\nabla _j$, it is
clear that the operator $H_A $ is  symmetric
on $C_0^\infty (\Omega)$.\\ 
In \cite {CT} we discuss the essential self-adjointness of this operator.
The  result in dimension $2$ is the following
\begin{theorem}\label{ess2}
Assume that $\pa \Omega $ is compact and that 
$B(x)$ satisfies near $\pa \Omega $
\begin{equation}\label{bune}
 {\bf b}(x)  \geq (D(x))^{-2}~, 
\end{equation}
then the Schr\"{o}dinger operator $ H_A$ is essentially self-adjoint.
($D(x)$ denotes the distance to the boundary). This  still holds true for  any gauge $A'$ such that $dA'=dA=B$.

\end{theorem}
We have, using Cauchy-Schwarz inequality,
 $$ | \langle   {\bf b}(x) u, u \rangle  | =
 | \langle [ \nabla _1 ,\nabla_2 ] u, u \rangle |
 \leq \|\nabla_1 u\|^2 + \|\nabla_2 u\| ^2 \quad u \in C_0^{\infty}(\Omega).$$ 
This gives the well-known lower bound
 \begin{equation}\label{H2} \forall u \in C_0^{\infty}(\Omega),~
h_A(u) \geq  \left|\int_{\Omega} {\bf b}(x)  |u|^2  |dx| \right|~.
\end{equation}
In this paper, we do not use the conditions (\ref{bune}) but we assume nevertheless that ${\bf b(x)}$ 
grows to infinity as $x$ approaches the boundary.  
 The operator  $ H_A^D$ defined by Friedrichs extension of the quadratic form $h_A $ 
 has a compact resolvent. By analogy with
magnetic bottles on the whole space (see \cite{AHS, Col1, tr}), such an operator is called a magnetic bottle on the disk. 

 We will deal with spectral estimates for the operator  $ H_A^D$, using a
 perturbative method: introducing an additional non-negative
 bounded and radial potential $V$, we obtain an upper bound of the number $N(A,V)$ of negative eigenvalues of the operator  $H_A^D -V$ 
(Theorem \ref{thin}) and deduce, for any $\lambda>0$, an upper bound of the number $N(H_A^D,\lambda)$ of  eigenvalues of the operator $H_A^D $ smaller than $\lambda$ (Theorem \ref{thine}).
Theorem \ref{thin}  can be seen as a magnetic version of the Cwikel-Lieb-Rosenblum inequality (see \cite{Cw,Li,Ro}).
The CLR inequality provides a bound on the number of negative eigenvalues of
  Schr\"odinger operators  in $\bR^d$ for $d \geq 3$ (without magnetic field) and is a particular case of Lieb-Thirring inequalities  (see \cite{LW,Lit}).

Eigenvalue bounds were recently studied for magnetic Hamiltonians on $\bR^2$, for constant magnetic fields  (see \cite{FO}), for  Aharonov-Bohm magnetic fields (see \cite{BEL, Lp})
 and for a large class of magnetic fields (see \cite{K}). However, in \cite{K}, the total magnetic flux $\phi= \frac{1}{2\pi}\int_{\bR^2 }{\bf b}(x) dx $
has to be finite and the dependence
on the  magnetic field  is not explicit even in the radial case. In our result, the total flux 
 is not necessarily finite (see example \ref{quabuc}) and the upper bound involves explicitly the square of the magnetic potential.\\ 
 Magnetic Lieb-Thirring inequalities  were also obtained for  Pauli operators  (see \cite{ESo,ESol}), and links  
 between magnetic and non-magnetic Lieb-Thirring inequalities were discussed in \cite{F}.

\subsection*{Aknowledgements} 

The author would like to thank A. Laptev for fruitful
discussions, Y. Colin de Verdi\`ere
for useful comments, J.P. Truc for the communication about Proposition \ref{jp} and the referee for careful reading and helpful suggestions. 

\section{Main results  }
\label{sec:contrex}

We consider a smooth magnetic field $B= {\bf b}(x) dx_1\wedge dx_2$ 
and a scalar potential $V$
on the unit disk
 $\Omega =\{x=(x_1,x_2)\in\bR^2|~ x_1^2+x_2^2=r^2<1\}$ so that

\begin{itemize} 
 \item 
 $(H_1)$\quad $ K= \inf_{x\in \Omega}{\bf b}(x)>0 $ and $ {\bf b}(x) \rightarrow +\infty$ as $D(x) \rightarrow 0$

(i.e as $x$  approaches  the boundary.)
\item  $(H_2)$ \quad $B$ is radially symmetric ( consequently we write $ {\bf b}(r) $ instead of ${\bf b}(x))$
\item $(H_3)$\quad $V \in L^1 (\Omega)$, $V$ radial and non-negative,
 $V$ bounded from above\ .
\end{itemize}
 From assumption $(H_1)$ and from inequality (\ref{H2}) we deduce  that for any gauge $A$ associated to $B$,
the operator $H_A^D$ has a compact resolvent, and
assumption $(H_3)$ entails that the negative spectrum of $H_A^D -V$ is discrete, where
 $H_A^D -V$ denotes
 the operator defined by Friedrichs extension 
of the quadratic form $h_A -V$.\\
Using  assumption $(H_2)$  we introduce polar coordinates  $ (r, \theta), (r \in \bR^+,\theta \in [0,2\pi[)$ and consider the 
following magnetic potential ~: \begin{equation}\label{vectpot}
  A = -a(r)\sin\theta dx_1 + a(r)\cos\theta dx_2 , \quad a(r) = \frac{1}{r} \int_0^r {\bf b}(t) t dt \ .
\end{equation} We have $\Di dA= B$ and
\begin{equation}\label{vectpote}
A= A(r) d\theta \quad {\rm with} \quad A(r)= r a(r) = \int_0^r {\bf b}(t) t dt \ .
 \end{equation}$A(r)$ is the flux of the magnetic field through the disk of radius $r$ centered in the origin. The function $a(r) =A(r)/r$ is well-defined (and smooth) at the origin and it is the amplitude of the magnetic potential $A$ in cartesian coordinates .\\
The first theorem provides an upper bound of the number $N(A,V)$ of negative eigenvalues of the operator  $H_A^D -V$ where $A$ is the magnetic
potential defined by (\ref{vectpote}).\\ {\it From now on, $A$ denotes this specific potential.}\\
Noticing that we have $N(A',V)=N(A,V)$ for any gauge $A'$ so that $dA'=dA=B$, we will prove the following

\begin{theorem}\label{thin}
If assumptions $ (H_1) (H_2) (H_3)$ are verified and  
if moreover  \begin{equation}\label{buni}
 {\bf b}(x)  \leq M (D(x))^{-\beta}~, \quad 0<\beta < \frac{3}{2}
\end{equation} for some $M >0$, 
then  $$N(A,V)\leq   \frac{1}{\sqrt{1-\alpha}}  \int_0^1 [(\frac{1}{\alpha} -1) \frac{A^2 (r)}{r^2} + V(r)]rdr + 2 \int_0^1  \left[ 1+|\log[r\sqrt K ]| \right] V(r) rdr\ 
$$
for any $\alpha \in ]0,1[$. \\This inequality still holds when we replace in the left-hand side  $N(A,V)$ by $N(A',V)$, where $A'$ is any gauge verifying $dA'=dA=B$.

\end{theorem}

The second theorem is a consequence of the first one and provides an explicit upper bound of the
number $N(H_A^D,\lambda)$ of the eigenvalues of $H_A^D$ 
smaller than any positive value $\lambda$~:
\begin{theorem}\label{thine}
If assumptions  $(H_1)$ and $ (H_2)$ are   verified and  if moreover 
$$ {\bf b}(x)  \leq M(D(x))^{-\beta}~, \quad 0<\beta < \frac{3}{2}$$for some  $M>0$, 
then  the number of  eigenvalues of the operator $H_{A}^D $ smaller than $\lambda$ satisfies, for any $\alpha \in ]0,1[$, the following inequality
 \begin{equation}\label{bunj}N(H_{A}^D,\lambda)\leq  c_ K\lambda+  \frac{\lambda}{ 2\sqrt{1-\alpha}} + \frac{\sqrt{1-\alpha}}{\alpha}  \int_0^1 \left(\frac{A(r)}{r}\right)^2 rdr 
\end{equation}
with \begin{itemize}\item $\quad \Di c_K = \frac{3-\log  K}{2}$ \quad \quad\quad \quad if \quad$ 0< K\leq 1$

\item $\quad\Di c_K =  \ \left[\frac{1+\log K}{2} +\frac{1}{ K}\right]$ \quad
if \quad $ K>1$,
\end{itemize} Inequality (\ref{bunj}) still holds when we replace in the left-hand side $N(H_{A}^D,\lambda)$ by $N(H_{A'}^D,\lambda)$, where $A'$ is any  gauge verifying $dA'=dA=B$.
\end{theorem}
\begin{remark}\label{mini} The minimum of  the right-hand side is obtained 
by taking  $$\alpha_\lambda = \frac{-3I + \sqrt{I^2+ 4I\lambda}}{\lambda -2I}$$
with $\Di I:= \int_0^1 \left(\frac{A(r)}{r}\right)^2  r dr $.\end{remark}

\begin{example}\label{quabuc}
 Consider a magnetic field $B$ as in the definition (\ref{quat}) below, and assume $b(r) \equiv 1$ and $\beta = 1$ . Then  $c_K = \frac{3}{2} $,  the chosen gauge is $  A(r)=\int_0^r {\bf b}(t) t dt= -\ln (1-r)-r  $  and 
  the corresponding value of $I$  is
\begin{equation}\label{quab} I = \int_0^1  \frac{[\ln (1-r)+ r ]^2}{r} dr =2\zeta(3) -\frac{3}{2}\ . 
\end{equation}
\end{example}

\section{Proofs}
\subsection{Proof of Theorem \ref{thin}}

Let us introduce the polar coordinates  $x= (r, \theta), r \in \bR^+,\theta \in [0,2\pi[
$. We have denoted by $A$ the following vector potential~:

\begin{equation}
A= A(r) d\theta \quad {\rm with} \quad A(r)= r a(r) = \int_0^r {\bf b}(t) t dt \ .
 \end{equation} Due to assumption (\ref{buni}) the magnetic field we  consider is
of the type\begin{equation}\label{quat} 
{\bf b}(r) = \frac{b(r)}{(1-r)^{\beta}}   \ ,\ {\rm with}\quad \max_{ [0, 1[} b(r) \leq M  \ \ 
 {\rm and} \quad\  \beta< \frac{3}{2}.\end{equation} 

We first prove the following
\begin{lemma} If $B$ satisfies (\ref{quat}), then
 we can find some constant $C$ so that $A$ writes $A = A(r) d\theta = r a(r) d\theta \ $ where
  \begin{itemize}
   \item  $\Di \mbox{if}\quad \beta < 1$\quad
$\Di   \max_{ [0, 1[ } a (r)
 \leq C  .$
 \item $\Di\mbox{if} \quad\beta = 1\quad 
\Di a(r) ={\tilde a} (r)\ln(1-r),\quad \mbox{with}\quad\max_{ [0, 1[ }{\tilde a} (r)
 \leq C  \ .$ 
 \item  $\Di \mbox{if}\quad \beta > 1\quad 
\Di a(r) =\frac{{\tilde a}(r)}{(1-r)^{\beta-1}},\quad \mbox{with}\quad \max_{ [0, 1[ }{\tilde a} (r)
 \leq C  .$\end{itemize}
In particular $\Di\int_0^1  \left(\frac{A (r)}{r}\right)^2 rdr <\infty\ .$

\end{lemma}
\begin{demo}
Let us explain the case $\beta > 1$. The method for the case $\beta =1$ is the same.\\
From (\ref{quat}) we get\\
$\Di 0\leq \frac{1}{r} \int_0^r {\bf b}(t) t dt\leq \frac{1}{r}\int_0^r b(t)t (1-t)^{-\beta} dt\leq M \int_0^r (1-t)^{-\beta} dt\leq M \frac{(1-r)^{-\beta +1}}{\beta-1}$
 and the result follows.\\
The case $\beta < 1$ is straightforward.
\end{demo}

\noindent We  come now to the proof of Theorem \ref{thin}, following the method of \cite{L}.
The quadratic form associated to $H_A^D -V$ can be rewritten as
\begin{equation}\label{quadr}
 h_{A,V} (u)= \int_0^1 \int_0^{2\pi} \left[|\frac{\pa u }{\pa r}|^2 - V(r) |u^2| + r^{-2} \left[ [\frac{\pa  }{\pa \theta } -iA(r)] u \right]^2 \right] r dr d\theta
\end{equation}
for any $\Di u \in C_0^\infty ([0,1[\times [0,2\pi[).$
Changing variables $r=e^t$ and denoting $w(t,\theta)= u(e^t,\theta)$ for $t \in ]-\infty, 0[$ and $\theta \in [0,2\pi[$ we transfer
the form $ h_{A,V} (u)$ to
\begin{equation}\label{qudr}
 {\tilde h}_{A,V} (w)= \int_{-\infty}^0 \int_0^{2\pi} \left[|\frac{\pa w }{\pa t}|^2 -  {\tilde V}(t) |w^2|+  \left[ [\frac{\pa  }{\pa \theta } -if(t)] w \right]^2 \right]  dt d\theta
\end{equation}
with 
$$ {\tilde V}(t)=  e^{2t} V(e^t),\quad f(t)=  A(e^t)\ .$$
By expanding a given function $w \in C_0^\infty ([-\infty,0[\times [0,2\pi[)$ into a Fourier series we obtain that
${\tilde h}_{A,V} (w)=\oplus_{l\in\bZ}   h_{A,V}^\ell (w_\ell) $
with
 $$ h_{A,V}^\ell(v)=\int_{-\infty}^0 |\frac{\pa v }{\pa t}|^2  + \left[ \left (\ell- f(t) \right )^2 -  {\tilde V}(t) \right] |v^2|\ dt\ ,    $$
and $w_\ell= \Pi_\ell (w)$ where $\Pi_\ell $ is the projector acting as
$$\Pi_\ell (w)(r,\theta)= \frac{1}{2\pi}\int_{0}^{2\pi}e^{il(\theta-\theta')} w(r,\theta')d\theta'\ .$$ 
We write, for any $\alpha \in ]0,1[$ and any $\ell\in \bZ^*$

$$ h_{A,V}^\ell(v) \geq  \int_{-\infty}^0 |\frac{\pa v }{\pa t}|^2  + \left[(1- \frac{1}{\alpha})f^2 (t) -  {\tilde V}(t) + (1- \alpha)\ell^2\right] |v^2|\ dt\ .   $$
Let us denote by $L_{\alpha}$ the operator associated via Friedrichs extension to the quadratic form 
 $$ q_{\alpha} (v)=\int_{-\infty}^0 |\frac{\pa v }{\pa t}|^2   +\left[(1- \frac{1}{\alpha})f^2 (t) -  {\tilde V}(t)\right] |v^2|\ dt\ .   $$
$L_{\alpha}$ and $q_{\alpha}$ depend on $V$ and $A$ but we skip the reference to $V$ and $A$ in notations for the sake of simplicity.
Since $$ h_{A,V}^\ell \geq q_\alpha + (1- \alpha)\ell^2\ , $$  the number $N( h_{A,V}^\ell )$ of negative eigenvalues of
$ h_{A,V}^\ell 
$ is smaller than the number of negative eigenvalues of $L_\alpha + (1- \alpha)\ell^2 $. Hence denoting by
$\{-\mu_k^{\alpha}\} $  
 the negative eigenvalues of $ L_{\alpha}$ 
and by  $I_\ell$ the set $    \{k \in \bN; -\mu_k^{\alpha}  + (1- \alpha) \ell^2 <0\}$ for any $\ell\in \bZ^*$, we get
$$N(A,V) \leq \sum_{\ell\in \bZ^*} \sum_{k\in I_\ell} 1 + N(h_{A,V}^0)\ .$$
Noticing that the sum in the right-hand side is taken over the $(\ell, k)$ so that $0< |\ell| \leq \frac{1}{\sqrt{1-\alpha}} \sqrt{\mu_k^{\alpha}}$
we write
\begin{equation}\label{nav} 
N(A,V) \leq \frac{2}{\sqrt{1-\alpha}} \sum_{k\in \bN} \sqrt{\mu_k^{\alpha}}+ N(h_{A,V}^0)\ .\end{equation}
Let us extend the functions $f$ and  ${\tilde V}$ to $\bR$ by  zero and denote respectively by $f_1$ and ${\tilde V_1}$ these 
extensions.\\
Since $C_0^\infty ([-\infty,0[)\subset C_0^\infty (\bR) $, the negative eigenvalues $\{-\nu_k^{\alpha}\}$ of  the operator  $ L_1^{\alpha}$ associated via Friedrichs extension to the quadratic form $$ q_1^{\alpha} (v)=
\int_{-\infty}^{+\infty} |\frac{\pa v }{\pa t}|^2   +\left[(1- \frac{1}{\alpha})f_1^2 (t) -  {\tilde V_1}(t)\right] |v^2|\ dt\    $$
 verify
\begin{equation}\label{navi}
 \sum_{k\in \bN} \sqrt{\mu_k^{\alpha}}\leq \sum_{k\in \bN} \sqrt{\nu_k^{\alpha}}.
\end{equation} 
Applying the sharp inequality of Hundertmarkt-Lieb-Thomas \cite{HLT} (see Appendix ) to the operator $ L_1^{\alpha}$ we get

$$\sum_{k\in \bN} \sqrt{\nu_k^{\alpha}} \leq \frac{1}{2} \int_{-\infty}^{+\infty} \left[( \frac{1}{\alpha}-1) f_1^2 (t)+ {\tilde V_1}(t)\right] dt$$
$$
\quad\quad\quad\leq \frac{1}{2}  \int_{-\infty}^{0} \left[(\frac{1}{\alpha}-1) f^2 (t)+  {\tilde V}(t)\right] dt$$
\begin{equation}\label{navii}\quad\quad\quad\leq  \frac{1}{2} \int_0^1  \left[( \frac{1}{\alpha}-1)\frac{A^2(r)}{r^2} + V(r)   \right] rdr\ . \end{equation}
To conclude we need the following 
\begin{lemma}\label{lmm} 
Assume that $K= \inf_{x\in \Omega}{\bf b}(x)>0 $.Then 
for any  $\varepsilon \in ]0,1[$
\begin{equation}\label{navaw}
 N(h_{A,V}^0)= N(h_{A,0}^0-V) \leq \frac {1}{\varepsilon}\int_0^1  \left[ 1+|\log (\sqrt{\frac {( 1-\varepsilon )K}{ \varepsilon }}r)| \right] V(r) rdr\ .
\end{equation}
In particular
\begin{equation}\label{navuw}
 N(h_{A,V}^0)\leq 2 \int_0^1  \left[ 1+|\log(\sqrt K r)| \right] V(r) rdr\ . 
\end{equation}
\end{lemma}
\begin{demo}
\begin{itemize}
\item Step 1~: From (\ref{H2}) we get that 
$\ h_A(u) \geq  K \int_{\Omega} |u|^2  |dx| \quad \forall u \in C_0^{\infty}(\Omega),$ 
which implies for $h_{A,0}^0$ (returning to the variable $r$ and considering $V\equiv 0$),
$$h_{A,0}^0 (w)=  \int_{0}^1  \left[|\frac{\pa w }{\pa r}|^2  +  r^{-2} A^2(r) |w^2| \right]  r dr$$
$$\quad \geq K \int_{0}^1   |w|^2  rdr \quad \forall w \in C_0^{\infty}([0,1])\ .$$ 
We write for any $\varepsilon \in ]0,1[$
\begin{equation}\label{nava}
 N(h_{A,0}^0-V) \leq N(\varepsilon h_{A,0}^0 +(1-\varepsilon)K -V)\leq  N\left( h_{A,0}^0 +\frac {( 1-\varepsilon )K}{ \varepsilon }-\frac {V}{\varepsilon}\right)\ ,
\end{equation}
where we have used the fact that multiplying an operator by a positive constant does not change the number of its negative eigenvalues.

\item Step 2~: We establish the following upper bound~:
\begin{equation}\label{navy}
 N(h_{A,0}^0+1-V) = N(h_{A,V}^0 + 1)\leq   \int_0^1  \left[ 1+|\log r| \right] V(r) rdr\ . 
\end{equation}
We have
$$h_{A,V}^0 (w)= \int_{0}^1  \left[|\frac{\pa w }{\pa r}|^2 + \left[ r^{-2} A^2(r)-   V(r) \right]  |w^2|  \right] r dr$$
$$\quad \geq 
\int_{0}^1  \left[|\frac{\pa w }{\pa r}|^2 -   V(r) |w^2|  \right] r dr\quad
 \forall w \in C_0^{\infty}([0,1])\ .$$
By the variational principle, 
\begin{equation}\label{navys}
 N(h_{A,V}^0 + 1)\leq N(P_0 +1 -V), 
\end{equation}
where $P_0$ is the operator generated by the closure, in $L^2([0,1],rdr)$ of the quadratic form
$$\int_{0}^1  |\frac{\pa w }{\pa r}|^2    r dr,\quad w \in C_0^{\infty}([0,1])\ .$$
Considering the mapping $U: L^2([0,1],rdr)\rightarrow L^2([0,1],dr)$ defined by $(Uf)(r)=r^{1/2} f(r)$
we get that
 \begin{equation}\label{navyc}
 N(P_0 +1 -V)\leq  N(T_0 +1 -V)
\end{equation}
where the operator $ T_0= U P_0 U^{-1}$ is the Sturm-Liouville operator on $L^2([0,1],dr)$
acting on its domain by
\begin{equation}\label{navyco}(T_0  u)(r) = -u"(r)-\frac{u(r)}{4r^2},\quad u(0)=u(1)=0 \ .\end{equation}
The upper bound (\ref{navy}) will follow from the properties of $G(r,r,1)$, the diagonal element of the integral kernel
 of $(T_0+ 1)^{-1}$. 
Precisely we have
\begin{equation}\label{navycoth} G(r,r,1) \leq   r( 1+ |\log r|), \quad r\in [0,1[\ .\end{equation}
The proof of (\ref{navycoth}) is given in Appendix .
The Birman-Schwinger principle then yields
\begin{equation}\label{navyci}
 N(T_0 +1 -V)\leq \int_0^1 G (r,r,1)V(r) dr \leq    \int_0^1  \left[ 1+|\log r| \right] V(r) rdr\ .
\end{equation}
This ends the proof of (\ref{navy}), together with the inequalities (\ref{navys}) and (\ref{navyc}).

\item Step 3~:  We mimick the previous method  to get, for any strictly positive number $k$
\begin{equation}\label{navw}
 N(h_{A,0}^0 + k^2 -V) \leq  \int_0^1  \left[ 1+|\log (kr)|  \right] V(r) rdr\ . 
\end{equation}
Due to the Birman-Schwinger principle it suffices to  prove that, for  any strictly positive number $k$   
\begin{equation}\label{navycwth} G(r,r,k^2) \leq   r( 1+ |\log (k r)| ), \quad r\in [0,1[\ .\end{equation}
This is done in Appendix .\\

\item Step 4~: Returning to  (\ref{nava}) and applying (\ref{navw}) with $k^2= \frac {( 1-\varepsilon )K}{ \varepsilon }$ and $\frac {V}{\varepsilon} $ instead of $V$
we get, for any  $\varepsilon \in ]0,1[$
\begin{equation}\label{navawe}
 N(h_{A,0}^0 -V) \leq  N \left( h_{A,0}^0 +\frac {( 1-\varepsilon )K}{ \varepsilon }-\frac {V}{\varepsilon}\right)\quad\quad\end{equation}
\begin{equation}\label{navaw}\quad\quad\quad \quad\quad\quad\quad\leq \frac {1}{\varepsilon}\int_0^1  \left[ 1+|\log (\sqrt{\frac {( 1-\varepsilon )K}{ \varepsilon }}r)| \right] V(r) rdr\ ,
\end{equation}
and taking $\varepsilon =\frac {1}{2}$ we obtain Lemma \ref{lmm}.
\end{itemize}\end{demo}

 Theorem \ref{thin} follows from Lemma \ref{lmm} together with inequalities (\ref{nav}), (\ref{navi}), and (\ref{navii}).

\subsection{Proof of Theorem \ref{thine}}

 Noticing that for any $\lambda >0$ the constant potential $V (x) \equiv \lambda$ is in $L^1 (\Omega)$, and that 
 $N(A,\lambda)$ denotes the number of eigenvalues of the operator $H_A^D $ less than $\lambda$, we apply Theorem \ref{thin}
to $V (x) \equiv \lambda$. To get the result it suffices to compute $\int_0^1  \left[ 1+|\log (kr)|  \right]  rdr\ $.
We get after computation that
\begin{equation}\label{navwz}
 \int_0^1  \left[ 1+|\log (kr)|  \right]  rdr\ = \gamma_k, \end{equation} with
\begin{itemize}\item \quad $\Di \gamma_k = \frac{3 - 2\log k}{4}$ \quad if $ k\leq 1$
\item \quad$\Di \gamma_k = \frac{1 + 2\log k}{4} +\frac{1}{2k^2}$ \quad
if $k>1$\ .
\end{itemize} 

\subsection{ Proof of Remark \ref{mini}}

 To get the minimum over the values of $\alpha$ we study the sign of the expression, for any $\alpha \in ]0,1[$, 
 of $$ g_{\lambda}(\alpha):=\frac{\lambda}{ 2\sqrt{1-\alpha}} + \frac{\sqrt{1-\alpha}}{\alpha}I\ . $$
A direct computation shows that the value $\alpha_{\lambda}$ which realizes the minimum of $g_{\lambda}(\alpha)$
 is the positive solution of
\begin{equation}\label{navite}
 \alpha^2 (\lambda -2I) + 6\alpha I -4I =0 \ .
\end{equation}
\section{An asymptotic eigenvalue upper bound}
From Theorem \ref{thine} we get easily an asymptotic estimate for the right-hand side of (\ref{bunj}) when $\lambda $ tends to $\infty$~:  
\begin{corollary}
If assumptions $(H_1)$ and $ (H_2)$ are  satisfied and  if moreover $$
 {\bf b}(x)  \leq M (D(x))^{-\beta}~, \quad 0<\beta < \frac{3}{2}
$$ for some $M>0$, 
then the number of  eigenvalues of the operator $H_A^D $ smaller than $\lambda$ satisfies, as $\lambda \rightarrow\infty$
\begin{equation}\label{quaba}
N(H_{A}^D,\lambda)\leq  ( \frac{1}{2}+ c_K) \lambda + \sqrt\lambda \sqrt I + O(1)\ , \end{equation}
where$$ I =\int_0^1 \left(\frac{A(r)}{r}\right)^2   rdr \ ,$$
and
 \begin{itemize}\item $\quad \Di c_K =  \frac{3-\log  K}{2}$ \quad \quad\quad \quad if \quad$ 0< K\leq 1$

\item $\quad\Di c_K =  \ \left[\frac{1+\log K}{2} +\frac{1}{ K}\right]$ \quad
if \quad $ K>1$\ .
\end{itemize}

 Inequality (\ref{quaba}) still holds when we replace in the left-hand side  $N(H_{A}^D,\lambda)$  by $N(H_{A'}^D,\lambda)$, where $A'$ is any  gauge  verifying $dA'=dA=B$.
\end{corollary}

\begin{demo}
We define as previously, for any $\alpha \in ]0,1[$, $$g_{\lambda}(\alpha):=\frac{\lambda}{ 2\sqrt{1-\alpha}} + \frac{\sqrt{1-\alpha}}{\alpha} I$$
and we want to determine the asymptotic behavior  as $\lambda $ tends to $\infty$ of $g_{\lambda}(\alpha_{\lambda})$, where  $\alpha_{\lambda}$ is the minimum of $g_{\lambda}(\alpha)$.\\
From (\ref{navite})
we compute  the following asymptotics
$$\alpha_{\lambda} = \frac{ 2 \sqrt I }{\sqrt\lambda} + O(\frac{1 }{\lambda})$$
$$\sqrt{1-\alpha_{\lambda}} = 1 - \frac{ \sqrt I }{\sqrt\lambda} + O(\frac{1 }{\lambda})\ ,$$
and this gives the result.\\
\end{demo}
\begin{remark}
The leading term in the estimate (\ref{quaba}) is of the same order than the leading term in the Weyl formula for the Dirichlet Laplacian 
(corresponding to the case $A\equiv 0$) in the unit disk.
\end{remark}
\section{Appendix }
\subsection{The inequality of Hundertmarkt-Lieb-Thomas}
 We recall the  sharp inequality of Hundertmarkt-Lieb-Thomas  \cite{HLT} 
\begin{theorem}
Let $$Lv(t) =-v"(t)-W(t)v(t),\quad W\geq 0 \quad W \in L^1(\bR)
$$ be defined in the sense of quadratic forms on $\bR$, and assume that the negative spectrum of $L$ is discrete.
Denote by  $\{-\nu_k, k\in \bN\} $ the negative eigenvalues of $L$.
Then $$\sum_{k\in \bN} \sqrt{\nu_k} \leq \frac{1}{2} \int_{-\infty}^{+\infty}   W(t) dt\ .$$
\end{theorem}
\subsection{The Green function $G (r,r',1)$ of the operator $T_0$.}
Let us  compute the diagonal element for the Green function $G (r,r',1)$ of
the operator $T_0$ defined by (\ref{navyco}).
$G (r,r',1)$ is the solution of 
\begin{equation}\label{navycot}\left((T_0 + 1) u\right)(r) = \delta_{r'}(r),\quad u(0)=u(1)=0 \ .\end{equation}
We have

$G (r,r',1) = A_1 u_1(r)+ A_2 u_2(r)  \quad r\leq r'$

$G (r,r',1) = B_1 u_1(r)+ B_2 u_2(r)  \quad r > r'\ ,$\\
where
 $u_1(r) =\sqrt r I_0 (r)$ and $u_2(r) =\sqrt r K_0 (r)$ are independent solutions
of the related homogeneous equation, ($I_0$ and $K_0$ are the modified Bessel functions).\\
The coefficients depend of $r'$ but we omit the indices for the sake of clarity. Due to the boundary conditions and to
the fact that the derivative (with respect to $ r$) of $G (r,r',1)$ has the
discontinuity in $r'$ of a Heaviside function, they satisfy~:
$$A_1 u_1(0)+ A_2 u_2(0)=0 \quad\quad B_1 u_1(1)+ B_2 u_2(1)=0$$
$$B_1- A_1= \frac{-u_2(r')}{W(r')}  \quad\quad B_2- A_2= \frac{u_1(r')}{W(r')}$$
where $W(r')$ is the value of the Wronskian of $u_1$ and $u_2$ taken at the point $r'$.\\
The first equation is always satisfied since $u_1(0)= u_2(0)=0$. Let us set $A_2 =0$. We have 
 $W(r')=  u'_1(r')u_2(r')- u_1(r') u'_2(r') = r' {\hat W}(r')$ where ${\hat W}(r')$ is the Wronskian of 
the modified Bessel functions $I_0$ and $K_0$. As $r'{\hat W}(r') =1$ (see \cite{AS}), we get after solving the above system,
 and doing $r=r'$ ~:
$$G (r,r,1) = u_1(r) \left[ -u_1(r)\frac{u_2(1)}{u_1(1)} + u_2(r)\right]$$
$$\quad = r I_0(r)
\left[-I_0(r)\frac{K_0(1)}{I_0(1)}  + K_0(r)\right]\ .$$
Using again the properties of the modified Bessel functions (see \cite{AS}) we can 
write $$G (r,r,1) \leq r I_0(r) K_0(r)\ .$$ 
The function $$g(r) = \frac{ I_0(r) K_0(r)}{ 1+ |\log r|}$$   has a limit at $r=0$ 
 equal to $1$ (see \cite{AS}), so  \begin{equation}\label{bess}c_0 = \max_{ [0, 1[} \frac{ I_0(r) K_0(r)}{ 1+ |\log r|}\ \end{equation}  exists and
$$G (r,r,1) \leq c_0 r( 1+ |\log r|), \quad r\in [0,1[\ .$$
Numerics suggest   that $g$ is decreasing on $[0,1]$, so that one
should have $c_0 =1$. In  next subsection, we give the  proof of this result, which can not be found to our
knowledge in the 
literature, and has been communicated to the author by J.P. Truc \cite{jptr}~:
\begin{proposition}\label{jp}
$\displaystyle \forall r ~\in ]0,1] : \frac{I_0(r)K_0(r)}{1-\log r } \leq 1$.
\end{proposition} 


\subsection{ Proof of Proposition \ref{jp} }
The modified  Bessel function $I_0$ can be written as 
\begin{equation}
	I_0(r)=\sum_{k=0}^{+\infty}\frac{(\frac{r^2}{4})^k}{k!^2}=1+\frac{r^2}{4}+...
\end{equation}
Therefore we have 
$$1 \leq I_0(r)\leq \sum_{k=0}^{+\infty}\frac{(\frac{r^2}{4})^k}{k!}=e^{\frac{r^2}{4}}$$
and
\begin{equation}\label{letoto}
\forall r \in ]0,1]~:~1 \leq I_0(r) \leq e^{\frac{1}{4}}\ .
\end{equation}
According to the expression of the modified  Bessel function $K_0$ 
\begin{equation}
	K_0(r)=-\Big( \log(r/2)+ \gamma \Big)I_0(r)+ \sum_{k=1}^{+\infty}\Big(\sum_{j=1}^k \frac{1}{j}\Big)\frac{(\frac{r^2}{4})^k}{k!^2}
\end{equation}
where $\gamma$ denotes the Euler constant,
we compute that
 \begin{equation}\label{ltt} K_0(r)I_0(r)-(1-\log r)=\delta(r)-1\ ,\end{equation}
where $\delta (r)$ denotes the following function : 
\begin{equation}
\delta(r)=(1-I_0(r)^2)\log r  -\Big( -\log 2+ \gamma \Big)I_0(r)^2+ I_0(r)\sum_{k=1}^{+\infty}\Big(\sum_{j=1}^k \frac{1}{j}\Big)\frac{(\frac{r^2}{4})^k}{k!^2}.
\end{equation}
Proposition \ref{jp} is then a straightforward consequence of  the following Lemma 

\begin{lemma} \label{latata}
$$\forall r \in ]0,1]~:~ \delta(r) \leq 1.$$
\end{lemma} 
\begin{demo}
The function $\delta(r)$ splits into 3 positive parts, which we study separately .
\begin{itemize}
\item An upper bound  for $ (1-I_0(r)^2)\log r $.\\
From (\ref{letoto}) we deduce $1-I_0(r)^2 \geq 1-e^{\frac{r^2}{2}}$, and : 
$$\forall r \in ]0,1]~:~0 \leq (1-I_0(r)^2)\log r \leq  \Big(e^{\frac{r^2}{2}}-1 \Big)(-\log r) \leq 0,11.$$
\item An upper bound  for $\Big( -\log 2+ \gamma \Big)I_0(r)^2$.\\
A straightforward computation gives
$-\gamma +\log 2 \leq 0.12$ so using that $I_0(r) \leq e^{\frac{1}{4}}$ we get
$$\Big( -\log 2+ \gamma \Big)I_0(r)^2\leq 0.16.$$
\item An upper bound  for $\displaystyle I_0(r)\sum_{k=1}^{+\infty}\Big(\sum_{j=1}^k \frac{1}{j}\Big)\frac{(\frac{r^2}{4})^k}{k!^2}$.\\ 
For $k \in \mathbb N^*$, we set $\displaystyle s_k=\sum_{j=1}^k \frac{1}{j}$. We have $s_1=1$ .
For $k \geq 2$, according to the inequality
$$\frac{1}{k}\leq \int_{k-1}^k \frac{dt}{t}=\log k - \log (k-1).$$
we get that:  
$$\sum_{j=2}^k \frac{1}{j} \leq \log k$$
and for any integer $k$ , $s_k \leq 1+\log k$. Thus 
$$\sum_{k=1}^{+\infty}\Big(\sum_{j=1}^k \frac{1}{j}\Big)\frac{(\frac{r^2}{4})^k}{k!^2}
\leq \sum_{k=1}^{+\infty} \Big(\frac{1+\log k}{k!}\Big)\frac{(\frac{r^2}{4})^k}{k!}.$$
Noticing that,  for any integer  $k \geq 1$ 
 $$0\leq\frac{1+\log k}{k!}\leq \frac{1+\log k}{k}\leq 1,$$
we can write, $\forall r \in ]0,1]$ : 
$$\sum_{k=1}^{+\infty}\Big(\sum_{j=1}^k \frac{1}{j}\Big)\frac{(\frac{r^2}{4})^k}{k!^2} \leq 
  \sum_{k=1}^{+\infty}\frac{(\frac{r^2}{4})^k}{k!}=e^{\frac{r^2}{4}}-1 \leq e^{\frac{1}{4}}-1.$$
 Finally we have, for any $r \in ]0,1]$ 
$$  I_0(r)\sum_{k=1}^{+\infty}\Big(\sum_{j=1}^k \frac{1}{j}\Big)\frac{(\frac{r^2}{4})^k}{k!^2} \leq e^{\frac{1}{4}}\Big(e^{\frac{1}{4}}-1\Big)  \simeq 0.364\ .$$ 
\end{itemize}
Summing the 3 previous estimates one gets   : 
$\Di\forall r \in ]0,1] ~:~\delta(r) \leq 0.11+0.16+0.37 \leq 1\ .$ \end{demo}\\
 The optimality of the value  $c_0=1$ is due to the fact that
$$\lim_{r\to 0+}\frac{K_0(r) I_0(r)}{1-\ln r}=1.$$

\subsection{The Green function $G (r,r',k^2)$ of the operator $T_0$}

We now compute the diagonal element for the Green function $G (r,r',k^2)$ of
the operator $T_0$ defined by (\ref{navyco}).
$G (r,r',k^2)$ is the solution of 
\begin{equation}\label{navycot}\left((T_0 + k^2) u\right)(r) = \delta_{r'}(r),\quad u(0)=u(1)=0 \ .\end{equation}
We have, as previously
$$G (r,r,k^2) = u_1(r) \left[ -u_1(r)\frac{u_2(1)}{u_1(1)} + u_2(r)\right]$$
where  $u_1(r) =\sqrt r I_0 (kr)$ and $u_2(r) =\sqrt r K_0 (kr)$ are independent solutions
of the related homogeneous equation.
This leads to
$$G (r,r,k^2) =r I_0(kr)
\left[-I_0(kr)\frac{K_0(k)}{I_0(k)}  + K_0(kr)\right]\leq r I_0(kr) K_0(kr) \leq  r( 1+ |\log (kr)| )\ .$$

\end{document}